\documentclass[10pt,twoside]{hsqcd}
\usepackage{epsf,epsfig,amsmath,amssymb,bm}
\usepackage{graphicx}
\usepackage{amssymb}

\begin{document}

\title{{\hfill RUB-TPII-09/08}\\ [1cm]Theory Summary of the Hadron
                                      Structure and QCD Workshop 2008
        \thanks{Theory summary talk presented at International
        Workshop ``Hadron Structure and QCD'' (HSQCD'2008),
        June 30 - July 4, 2008, Gatchina, Russia}
      }

\author{\underline{N.~G.~STEFANIS} \\ \\
Institut f\"ur Theoretische Physik II \\
Ruhr-Universit\"at Bochum \\
D-44780 Bochum, Germany \\
E-mail: stefanis@tp2.ruhr-uni-bochum.de }

\maketitle

\begin{abstract}
\noindent
This paper is the write-up version of the theory summary talk given at
the HSQCD 2008 Workshop in Gatchina, Russia.
Recent theoretical developments and results are summarized focusing on
works that point out new perspectives, ranging from perturbative QCD,
DGLAP and ERBL evolution, polarized and unpolarized parton distribution
functions, small-x physics to nonperturbative QCD, lattice
simulations, quark-gluon matter, hadron spectroscopy, and predictions
for the LHC.

\end{abstract}



\markboth{\large \sl \underline{N.~G.~STEFANIS}
\hspace*{2cm} HSQCD 2008} {\large \sl \hspace*{1cm} Theory Summary}

\section{Introduction}
\label{sec:intro}

The Workshop on the Hadron Structure and QCD 2008: From Low to High
Energies continues the tradition of a series, started in 2004.
This year's meeting provided a
\begin{itemize}
\item balanced mixture of theoretical and experimental talks,
\item high level of contributions,
\item good mixture of experienced and young scientists,
\item rich social program.
\end{itemize}
The aim of the workshop was to review the progress in hadronic physics,
QCD, and the Standard Model and its generalizations, focusing on:
\begin{itemize}
\item
perturbative QCD, BFKL- and DGLAP- evolution,
\item
polarized and unpolarized parton distribution functions,
\item
small-x physics,
\item
hard diffraction and Pomeron physics,
\item
heavy-ion collisions and quark-gluon matter,
\item
nonperturbative QCD, lattice computations, and chiral models of hadrons,
\item
hadron spectroscopy and exotic states,
\item
precision tests of the Standard Model,
\item
extensions of the Standard Model and
predictions for the Large Hadron Collider (LHC) (and future colliders).
\end{itemize}

This theory summary is set in the broader context of hadronic physics
and in the narrower context of QCD.
Within this context, various theoretical papers---out of a total of
fifty--have been selected for a more detailed presentation, weaving them
into a theoretical frame in such a way as to convey the coherence of
the underlying ideas and methods.
The guide and motivation are the wide range of possible applications
and their relevance in pursuing future developments in the advent of
the LHC at CERN.
These theoretical contributions are supplemented by thirty experimental
talks, summarized by M.~Sapozhnikov.
However, it is outside the scope of this summary to provide rigorous
and critical assessments on the pertinence and validity of the
presented material.
The theoretical contributions to follow are grouped by broad categories
of concern:
QCD Calculations (Sec.\ \ref{sec:QCD-calc}),
Evolution Equations and Related Topics (Sec.\ \ref{sec:evolution}),
QCD and Higgs Physics (Sec.\ \ref{sec:Higgs}),
Regge Physics and Diffraction (Sec.\ \ref{sec:regge}),
Nonperturbative QCD and Low-energy Models (Sec.\ \ref{sec:npQCD}),
Hadron Form Factors (Sec.\ \ref{sec:form-factors}),
Quark Confinement (Sec.\ \ref{sec:confin}), and
Mathematical and Other Analyses (Sec.\ \ref{sec:math}).
I hope that readers of these proceedings will find, whether or not
they agree with particular arguments, that the presented contributions
are in general competent and useful.

\section{QCD Calculations}
\label{sec:QCD-calc}

Parton distribution functions (PDF)s---unpolarized and polarized---give
the probability to find partons (quarks and gluons) without and with
spin in a hadron as a function of the longitudinal momentum fraction,
$x$, carried by the parton, and the hard scale $Q^2$ (see \cite{Sop96}
for an introduction).
All other components of parton momentum are integrated over.
These integrated PDFs have in the $\overline{\rm MS}$ scheme
unambiguous gauge-invariant definitions in terms of matrix elements of
operators and satisfy factorization theorems \cite{CS82}---see, e.g.,
\cite{CSS89} for a review.
When the integral over the parton's transverse momentum is not
carried out, one deals with unintegrated, or transverse-momentum
dependent (TMD), PDFs whose gauge-invariant formulation is still an
issue under scrutiny (see Stefanis' talk).

The spin structure function $g_1$ at low $x$ and
arbitrary $Q^2$, as measured by the COMPASS Collaboration at CERN, was
considered in Greco's talk \cite{EGT08}.
The hadronic tensor $W_{\mu\nu}$ contains a symmetric part that does
not depend on spin and an antisymmetric one, parameterized by two
structure functions, viz., $g_1$ and $g_2$, which depend on $Q^2$ and
$x = Q^2/2p\cdot q$ $(0< x < 1)$.
These quantities involve both perturbative and nonperturbative QCD
ingredients and are, therefore, model-dependent.
On account of factorization, one can cast $W_{\mu\nu}$ in the form of
a convolution which separates out the nonperturbative content into the
probabilities $\Phi_{\rm quark}$ and $\Phi_{\rm gluon}$.
These functions (called the initial quark and gluon densities and
denoted by $\delta q$ and $\delta g$) cannot be calculated from first
principles of QCD and have to be either modeled or fitted to
experimental data at large $x\sim 1$ and large $Q^2$.
Moreover, each structure function has both a non-singlet and a singlet
component:
$g_1 = g_1^{\rm NS} + g_1^{\rm S}$
(analogously for $g_2$).
Recall that there are various kinematical regions to cover in the
($1/x~,Q^2$) plane.
For instance, one has for $x\leq 1$ and $Q^2\gg \mu^2$ the DGLAP region,
whereas for $x\ll 1$ and $Q^2\leq \mu^2$ one probes the COMPASS region.
Then, one can use the perturbative DGLAP $Q^2$- evolution
equation---together with fits for the initial parton densities---to
predict $g_1^{\rm NS/S}(x,Q^2)$.

However, though DGLAP evolution resums all terms
$\sim \left(\alpha_s\ln(Q^2/\mu^2)\right)^{k}$, it does not account
for the resummation of logarithms of $x$.
For that reason, Greco and his coauthors have suggested another
resummation procedure of the logarithms of $1/x$ which employs an
infrared cutoff in the transverse space $(k_\perp^2 > \mu^2)$.
This gives rise to the evolution of the structure functions with respect
to the variable $\mu^2$ (a method employed before by several authors,
including Gribov, Lipatov, Kirschner, Bartels, Ermolaev, Manaenkov, and
Ryskin).
This Infra-Red Evolution Equations (IREE) assume for $g_1^{\rm NS}$
a form which is driven by an anomalous dimension summing up all
double and some single logarithms of $x$.
In contrast to DGLAP evolution which is reliable at large $x$, but
fails at small $x$ lacking the total resummation of logarithms of $x$,
the discussed approach is good at small $x$, turning bad at large
$x$, because it neglects some contributions essential in this region.
To merge both approaches, the authors involve an IR cutoff scale $\mu^2$
and shift $x \to  \bar{x} = (Q^2 + \mu^2)/2p\cdot q$ already at the
level of the involved Feynman graphs.
This amounts to introducing a ''mass'' of virtual quarks and gluons
to regulate IR singularities, i.e.,
$
  \int_{\mu^2}^{Q^2}\frac{dk_{\perp}^{2}}{k_{\perp}^{2}}
\to
  \int_{0}^{Q^2}\frac{dk_{\perp}^{2}}{k_{\perp}^{2}+\mu^2}
=
  \ln \left( \frac{Q^2 + \mu^2}{\mu^2} \right) \ .
$
[We will encounter another IR protection within the framework of
Fractional Analytic Perturbation Theory (FAPT) in Bakulev's talk.]
Using the scheme described, Greco {\it et al.} obtain the following
results:
(i) They predict that at small $Q^2$, $g_1$ is almost independent of
$x$, even at $x \ll 1$, and depends only on $2p\cdot q=w$.
(ii) They suggest that it would be interesting to study the
$w$-dependence of $g_1$ at COMPASS.
(iii) This would allow one to answer the question about the sign of the
gluon density and fix the ratio $N_g/N_q$.
(iv) The report concluded that using another target would allow to
measure the non-singlet distribution $g_1$ and define $N_q$.

I presented this contribution in some technical and conceptual detail,
because it contains several ingredients that appear also in other
papers presented in this workshop.

Transverse-momentum dependent PDFs were considered also by Teryaev.
He studied the relations between leading and higher twists in
nonperturbative QCD in terms of matrix elements of quark/gluon
operators as resummed towers of twists.
In particular, he addressed the transverse moment of the Sivers
function \cite{Siv90} in the context of Single-Spin Asymmetries (SSA)
\cite{Col02,BJY02,BHS02,BMP03}.
He pointed out that TMD PDFs of (naively) leading twist may
turn into an infinite sum of higher twists.
In the case of the Sivers function, this issue may be assessed by means
of the T-invariance or, technically speaking, by considering the
imaginary part of the (quark) density matrix.
An important finding here is that the Sivers function appears to be
related to the twist-3 gluonic poles.

In the talk of Stefanis the anomalous dimensions of fully
gauge-invariant TMD PDFs \cite{JY02,BJY02} in the light-cone
gauge were considered.
It was shown in \cite{CS07,CS08}, and reported in these proceedings
\cite{SC08_HSQCD08}, that associating individual gauge contours of
integration to the quark field operators in the quark-pair correlator
describing the distribution of a quark in a quark, additional
ultraviolet (UV) divergences appear.
The origin of these UV divergences was found to be rooted in the
renormalization effect on the junction point of Wilson lines, when
they contain transverse segments extending to light-cone infinity.
An explicit one-loop calculation in the light-cone gauge $A^+=0$ shows
that the anomalous dimension ensuing from these divergences coincides
with the one-loop expression of the universal cusp anomalous dimension
\cite{KR87}.
To dispense with this anomalous-dimension defect and obtain the same
result for the renormalized TMD PDFs as in a covariant gauge (say, the
Feynman gauge), a modified definition of the TMD PDFs was proposed.
This contains a soft counter term in the sense of Collins and Hautmann
\cite{CH00} which is a path-ordered exponential factor evaluated along
a particular gauge contour with a cusp.
This integration contour stretches out to light-cone infinity and
ventures off the light cone in the transverse direction.
The anomalous dimension associated with the renormalization of this
nonlocal operator compensates the anomalous-dimension artifact and
ensures that integrating over the parton transverse momentum, one finds
a PDF satisfying the DGLAP evolution equation.
Moreover, the anomalous dimension of the modified TMD PDF respects the
Slavnov-Taylor identities and resembles the one-loop expression one
finds for a TMD PDF with a connector \cite{Ste83} (see also
\cite{CD80}) insertion, i.e., the direct Wilson line between the two
quark fields.
The cusp-like junction point is ``concealed'' by light-cone infinity,
and reveals itself only \textit{after} renormalization as a phase
entanglement \cite{CS08} akin to the ``intrinsic'' Coulomb phase,
found before in QED \cite{JS90}, and being codified in the (one-loop)
cusp anomalous dimension.
The implications of a more accurate definition of TMD PDFs are far
reaching, ranging from more precise analyses of various experimental
data on hard-scattering cross sections to the development of more
accurate Monte Carlo event generators.

\section{Evolution equations and related topics}
\label{sec:evolution}

Wilson lines were also on the focus of Balitsky's presentation.
He gave a status report on the next-to-leading order (NLO)
evolution of color dipoles (i.e., a two-Wilson-lines operator),
exploiting the NLO evolution kernel in detail \cite{BC08}.
This kernel consists of three parts:
(i) a running-coupling part proportional to the $\beta$ function,
(ii) a conformal part describing a $1\to3$ dipole transition, and
(iii) a non-conformal part.
The author provided evidence that the result agrees with the forward
NLO BFKL kernel up to a term proportional to $\alpha_{s}^{2}\zeta(3)$
(where $\zeta(s)$ is Riemann's zeta function) times the original
dipole.
Moreover, he argued that for the creation of dipoles in the small-$x$
evolution, the argument of the coupling constant is determined by
the size of the smallest dipole \cite{BC08}.
It turns out that with a rigid $|\alpha_s|<\sigma$ cutoff, the NLO BK
kernel in $N=4$ supersymmetric Yang-Mills (SYM) theory is (almost)
conformally invariant in the transverse plane.

From Lipatov's talk we learned about new calculations of scattering
amplitudes in $N=4$ SYM theory with BFKL kernels at the two-loop order.
In this approach, the Pomeron is a composite state of Reggeized gluons,
for which an effective action was derived.
The integrability of the equations for the multi-gluon states was
proved and it was shown that the BFKL dynamics is integrable in Leading
Logarithmic Approximation (LLA).
A supplementary analysis was presented by Hentschinski.
He reported on the computation of longitudinal loop integrals in a
gauge-invariant effective action for high-energy QCD \cite{HBL08}.
He described how longitudinal integrations up to two $t$-channel gluons
should be performed using the effective action.
Finally, he demonstrated that the obtained result for the elastic and
the production amplitude reproduces correctly the leading logarithmic
contribution and the energy discontinuities.

Saleev's talk (see also Shipilova's talk) was devoted to an
investigation of DIS and the prompt photon production in the Regge
limit of QCD in terms of Reggeized high-energy amplitudes (or,
equivalently, of Reggeized quarks and gluons) and effective vertices.
Using the QMRK (quasi-multi-Regge kinematics) approach---developed
by teams from St. Petersburg and Novosibirsk---and the Reggeized
quark ansatz, explicit analytic expressions for
$F_{2}(x_{\rm B},Q^2)$ and $F_{\rm L}(x_{\rm B},Q^2)$ were obtained.
Recall in this context that in the case of Reggeized gluons, Feynman
rules for the effective theory have been derived on the basis of
Lipatov's non-Abelian gauge-invariant action \cite{Lip95}.
More recently, Antonov, Kuraev, Lipatov, and Cherednikov \cite{ALKC04},
derived the Feynman rules directly from the effective Reggeon-particle
action and computed explicit expressions for some important effective
vertices.
Saleev and collaborators obtained predictions for $F_2$, $F_L$ by
employing LO quasi-multi-Regge kinematics and Kimber-Martin-Ryskin
quark (and gluon) unintegrated PDFs \cite{KMR01}.
They found that these functions are consistent with the prompt-photon
production data measured at the Tevatron.
In particular, agreement with the D$\emptyset$ and the CDF data for prompt
photons was found by considering $Q\overline{Q}\to\gamma$ as the main
production mechanism.
Using the QMRK approach, Shipilova and Saleev investigated the D-meson
production, measured at the Tevatron and at HERA.
Shipilova presented in her talk calculations of the $p_T$-spectra of
the $D$-meson photoproduction at HERA and found satisfactory agreement
at high $p_T$ between their predictions for the production (via LO QMRK)
of the subprocess $\gamma^* Q \to q$ (where $Q$ denotes the Reggeized
quark) and the experimental data.
Remarkably, the calculated $p_T$-spectra of the $D$-meson production at
the Tevatron via two LO QMRK processes also show agreement
with the experimental data.
In view of this outcome, she concluded that the unintegrated $c$-quark
and gluon distribution functions in the proton seem to be correct,
given that they yield good agreement between theory and the data for
different reactions.

In Ermolaev's contribution (in collaboration with Troyan) the
parametrization of the QCD coupling in evolution equations---including
DGLAP---was studied \cite{ET08_HSQCD08,ET08}, with particular attention
payed to the appropriate scale setting for the argument of the QCD
coupling.
The presented analysis of the parametrization of $\alpha_s$ pertains to
a wide group of evolution equations of the Bethe--Salpeter type,
including BFKL and DGLAP, where one virtual gluon is factorized out of
the blob.
Such a gluon can propagate either in the $s$-channel or in the crossing
channels and the parametrization of $\alpha_s(\mu^2)$ depends on the
considered channel.
In the $s$-channel, an effective coupling was derived that incorporates
$\pi^2$-terms due to the analytic continuation from the Euclidean to the
Minkowski region (cf.\ Bakulev's presentation).
For large values of $\mu^2$, these contributions can be neglected and
one recovers the standard DGLAP $\alpha_s$-parametrization.

\section{QCD and Higgs physics}
\label{sec:Higgs}

While the QCD strong coupling at high momenta (energies) is
small---thanks to asymptotic freedom---its low-momentum behavior
cannot be controlled by perturbation theory.
In fact, at Euclidean momenta $Q^2\sim \Lambda_{\rm QCD}^2$, the
one-loop $\alpha_s(Q^2)$ exhibits a Landau pole which is purely
unphysical.
As a result, the analytic continuation of the standard strong coupling
to Minkowski space fails.
Various proposals have been suggested over the past three decades how
to avert the Landau singularity of $\alpha_s$ at one as well as at
higher loops in the spacelike as well as how to define it in the
timelike region.
A crucial step forwards represents the so-called analytic perturbation
theory (APT), initiated by Shirkov and Solovtsov \cite{SS97} and
recently reviewed in \cite{SS06}.
Underlying this approach is the spectral representation of the strong
coupling in the Euclidean region in terms of a universal spectral
density which allows to define---under the proviso of
renormalization-group invariance---an analytic coupling simultaneously
in the Euclidean and in the Minkowski space.
Also applications to the ultra-low momentum region have been carried
out recently \cite{BNPSS07}, and alternative formulations of the
strong coupling below the Landau ghost singularities have been proposed
with the goal to include nonperturbative input \cite{NP04,CV07}.
A generalization and conceptual extension of APT was developed in a
series of works during the last decade, starting with applications to
the calculation of the factorizable part of the pion's electromagnetic
form factor in QCD \cite{SSK98} which typifies exclusive processes.
This study was continued and refined in \cite{BPSS04}.
In the year 2001 Karanikas and Stefanis generalized the analyticity
imperative by demanding that all terms in a QCD amplitude that can
affect the discontinuity across the cut along the negative real axis
$-\infty < Q^2 < 0$, and hence contribute to the spectral density,
have to be included into the analytization procedure, i.e., into the
dispersion relation \cite{KS01,Ste02}.
This work paved the way for further extending the whole analytic
approach beyond the level of integer powers of the coupling and
find analytic images of any real power in both the Euclidean
\cite{BMS05} and in the Minkowski region \cite{BMS06}, finally
culminating into the creation of Fractional Analytic Perturbation
Theory.
Using again the spacelike electromagnetic form factor as a
``theoretical laboratory'', it was shown in \cite{BKS05} (see also
\cite{Ste04,SK08,Bak08})
that FAPT provides a diminished sensitivity of the predictions on all
perturbative scales---the renormalization as well as the factorization
scale---while reducing significantly also the dependence on the
renormalization-scale setting procedure (and scheme) used.
In the Minkowski region, where Landau ghosts are absent, the
application of FAPT \cite{BMS06} to the decay of a scalar Higgs into
a $\bar{b}b$ pair at the four-loop level has provided expressions that
incorporate all $\pi^2$ terms induced by analytic continuation.

Using the FAPT methodology, Bakulev raised in his talk the important
question as to what order of the perturbative expansion one has to go
in order to find an estimate, say, for the Higgs boson decay, with an
acceptable precision (to parallel experimental results).
In fact, Bakulev showed that both APT and FAPT produce finite resummed
answers for perturbative quantities, provided one knows the generating
function for the coefficients of the perturbative-series expansion.
Recall that within the analytic approach one deals not with a power
series, but with functional non-power-series expansions.
Using a simple model for the generating function \cite{BM08} pertaining
to the Higgs-boson decay $H\to\overline{b}b$, it was concluded that at
the N$^3$LO the obtained accuracy of the truncation is of the order
of 1\%.
On the other hand, for the Adler function $\mathcal D(Q^2)$, an
accuracy of the order of as high as 0.1\% was reached already at
N$^2$LO (for more details, see \cite{Bak08_HSQCD08}).
These are encouraging results for further applications, given the
extreme complexity and computational amount of work in computing
high-order corrections in standard QCD perturbation theory.

Staying within the same subject, let us continue with Kim's talk which
considered the calculation of the main Higgs-boson decay width
into bottom quarks, and the role of higher-order QCD corrections and
their resummation.
Different methods for treating the results of higher-order perturbative
QCD calculations of this quantity were examined and their outcomes
compared.
Special attention was paid to the analysis of the dependence of the
decay width on the Higgs mass in the cases when the $b$-quark mass is
defined as the running parameter in the $\overline{\rm MS}$ scheme and
as the quark pole.
An important observation was that the results obtained with different
methods yield effects of $O(\alpha_s)$-corrections that are consistent
to each other.
This applies in particular to the estimated theoretical precision of
these results with respect to $\Gamma_{H\to\overline{b}b}$.
Furthermore, Kim discussed a means of verifying the stability of the
results against the inclusion of higher-order effects up to
$\alpha_s^4$, calculated in \cite{BCK06}.\footnote{Taka Yasuda reported
about Higgs searches at the Tevatron on behalf of the D$\emptyset$ and
CDF collaborations at Fermilab.}
He pointed out that the obtained predictions match those extracted
from FAPT without \cite{BMS06,SK08} and with the use of resummation
techniques \cite{BM08}.

Higgs production was also the subject of Strikman's talk.
He considered hard processes in high-energy $pp$ scattering as an
important tool in the search for new heavy particles---in particular of
the Higgs at LHC---via a diffractive process in which the produced
heavy particle is separated from the projectile fragments
by large rapidity gaps \cite{FHSW06}.
A gap survival in the mean field approximation (i.e., when there is no
correlation between hard and soft interactions in the impact parameter)
was considered with a possible strong suppression of this effect due to
the onset of the so-called black disk regime.
[The name derives from the fact that at high energies, strong
interactions enter a regime in which cross sections are comparable to
the ``geometric size'' of the hadrons and unitarity becomes
an essential feature of the dynamics.]
A crucial observation here is that the transverse area occupied by
partons with $x > 0.05$ is much smaller than the transverse area
associated with the proton in soft interactions.
This is, because of color fluctuations in fast nucleons and the slow
space-time evolution of their wave function.
Strikman provided evidence that the gap survival probability at
the LHC should be much smaller ($< 0.01$)---owing to the onset of the
black disk regime (or regime of high gluon field)---as compared to
models which neglect correlations of partons in the transverse plane.
A safe contribution was found to come from the region with
$b>1.2$~fm, leading to $S^2\geq 0.004$.
Hence, the $t$-dependence may provide a critical test in
distinguishing different mechanisms for the rapidity-gap suppression.

Another exciting application of Higgs physics was discussed by Khoze,
referring to a recent published work \cite{HKRSTW07}.
In his talk he assessed the Higgs sector beyond the Standard Model
(SM), suggesting forward proton tagging at the LHC.
His main aim was to demonstrate that the Central Exclusive Diffractive
Production can provide unique advantages for probing the non-SM Higgs
sector.
Indeed, the Forward Proton Tagging (FPT) would significantly extend the
physics reach of the ATLAS and CMS detectors by giving access to a wide
range of various channels pertaining to new physics effects.
Khoze underlined that FPT has the unique potential to enable such
measurements at the LHC---even being able to challenge those at the
International Linear Collider (ILC).
It turns out that for certain non-Standard-Model scenarios, FPT may
become the Higgs discovery channel at all, remarked Khoze, offering a
sensitive probe of the CP structure of the Higgs sector.

The possibility of soft diffraction at the LHC was addressed by
Ryskin (these Proceedings \cite{KMR_HSQCD08} and \cite{RMK07,MKR05}).
Two recent global analyses of available soft data, taken at the
CERN-ISR (Intersecting Storage Ring) up to the Tevatron energy range,
could be reproduced.
Ryskin stressed that the large bare triple-Pomeron (denoted below by
the label $\mathbb{P}$) coupling regime, described by
$g_{3\mathbb{P}}\sim 0.25 g_N$ (where $g_N$ is the nucleon-Pomeron
coupling) and $\Delta = \alpha_\mathbb{P}(0)-1\sim0.3$
(with $\alpha_\mathbb{P}(0)$ being the `bare Pomeron' intercept),
could predict $\sigma_\text{tot}\sim 90$~mb at LHC by taking recourse
to screening effects due to the soft$\leftrightarrow$hard Pomeron
transition.
The exclusive central diffractive production, $pp\to p+A+p$, has great
advantages for studying the Higgs sector at the LHC, because it would
allow to measure the mass of the Higgs boson with a very high
resolution \cite{MKR05}.
However, the expected number of events in the SM case is expected to be
rather small.
Extending the study to SUSY Higgs bosons, there are regions of the SUSY
parameter space, where the signal could be enhanced by a factor of 10
or more, while the background remains unaltered.
This opens up the possibility to discover new physics at the Tevatron
(as well as at the LHC).

An important question in the context of the exclusive central
diffractive production of the $b\bar{b}$ cross section at the LHC
concerns the size of QCD radiative corrections.
An in-depth analysis of this issue was presented by Shuvaev, who
reported about a recent work published in \cite{SKMR08}.
The amplitude for the $gg\to\overline{b}b$ production amplitude was
calculated for a color-singlet $J_z=0$ digluon state at
$O(\alpha_s^2)$.
It turns out that the radiative QCD (one-loop) corrections were found
to suppress the exclusive $\overline{b}b$ background by a factor
$\sim 2$ (or more for larger $\overline{b}b$-masses) for the central
exclusive diffractive Higgs production, in comparison with calculations
using the Born $gg\to\overline{b}b$ amplitude.

\section{Regge Physics and Diffraction}
\label{sec:regge}

Nikolaev reported on an extensive study \cite{NS06a,NS06b} of
multi-Pomeron vertices in QCD using nonlinear $k_{\perp}$
factorization.
He made the following observations:
(i) The concept of a coherent (collective) nuclear glue proves
extremely useful for the formulation of the Reggeon field theory
vertices of multi-Pomeron—-cut and uncut—-couplings to particles and
between themselves.
(ii) The concept of the collective nuclear glue as a coherent state of
the in-vacuum (Reggeized) gluons provides a useful tool, with the
nuclear collective glue defining an observable by coherent diffraction.
(iii) Nonlinear $k_{\perp}$-factorization quadratures for hard
scattering off nuclei with a fixed multiplicity of color-excited
nucleons could be derived and an expansion of nuclear unintegrated glue
in terms of the collective glue of overlapping nucleons and coherent
nuclear gluons could be performed.
(iv) It was demonstrated how the coupled-channel non-Abelian
intranuclear evolution of color dipoles, inherent in perturbative QCD,
gives rise to the Reggeon field theory diagrams for the final states.
(v) Remarkably, the coherent diffraction does not factorize into the
photon impact factor and the triple-Pomeron vertex.

It is clear from the previous report that $k_{\perp}$-factorization
is a powerful theoretical tool that can be used to analyze various
processes.
In Sch\"{a}fer's talk \cite{Sch08} unitarity cutting rules for hard
processes on nuclear targets were discussed \cite{NS06a} and some
applications for DIS off heavy nuclei were studied \cite{NSZZ06}.
It was found that topological cross-sections follow directly from the
nonlinear $k_{\perp}$-factorization for the inclusive cross-sections.
A novel property of the QCD unitarity cutting rules is that it gives
rise to two kinds of cut Pomerons.
It turns out that the topological cross-sections in DIS are
substantially different from the naive application of the
Glauber--Abramovsky, Gribov, and Kancheli (AGK) \cite{AGK73} approach
for color dipoles.

\section{Non-perturbative QCD and Low-Energy Models}
\label{sec:npQCD}

Recent experimental measurements of the anomalous magnetic moment of
the muon, $a_\mu = (g_\mu - 2)/2$, by the E821 experiment at the
Brookhaven National Laboratory have provided the value
$
  a_{\rm exp}
=
  11 659 208.0(6.3) \cdot 10^{-10} .
$
This unprecedented accuracy poses a challenge to theory within the SM
and asks for a theoretical uncertainty lower than the uncertainties
for the nearest-future experiments (at BNL, JPARC, and FNAL)
in order to be able to reveal effects of contributions beyond the SM.
Within the SM, one has
$
  a_{\mu}^{\rm SM}
=
  a_{\mu}^{\rm  QED} + a_{\mu}^{\rm EW} + a_{\mu}^{\rm Hadr}
=
  11 659 178.5(6.3)\cdot 10^{-10},
$
a prediction $3.4\sigma$ below the experimental value.
Because both $a_{\mu}^{\rm  QED}$ and $a_{\mu}^{\rm EW}$ are known
with a high accuracy, the only source of uncertainty (within the SM) is
the hadronic part, encoded in $a_{\mu}^{\rm Hadr}$.
The analysis \cite{DB08} presented by Dorokhov is devoted to the
calculation of the pion-pole contribution of the hadronic
light-by-light (LbL) scattering to $a_\mu$ within the nonlocal chiral
quark model (N$\chi$QM) \cite{ADT00,DB03}---motivated by the instanton
model of the QCD vacuum \cite{CDG77,Shu81,DP85}.
The benchmarks of the work include
a new estimate $a_{\mu}^{\text{LbL},\pi^0}=6.5\cdot10^{-10}$
which agrees with previous ones based on the usage of the CLEO data on
the pion-photon transition form factor.
Moreover, the QCD constraints suggested by Melnikov and Vainshtein
\cite{MV03} are satisfied within this model calculation.
The more demanding calculational task of including the scalar,
axial-vector, and the $\eta, \eta^{\prime}$ meson exchanges, as well as
taking into account the quark and meson box diagrams, is still in
progress.
\footnote{See also Eric Bartos' talk
on the pion-pole contributions to the LbL part of
$(g-2)_{\mu}$ \cite{BDDKZ01}.}
Hence, it remains to be seen whether the SM can fully explain the
measured value of $a_\nu$, or if there is room for new physics.

A new application \cite{GIS05} of Chiral Perturbation Theory (ChPT) to
pion polarizabilities was reported by Ivanov.
More specifically, he discussed a chiral expansion applied to the
$\gamma\gamma\to\pi\pi$
amplitude at the Compton threshold and found that the outcome
converges quite rapidly.
The obtained two-loop result for the charged pion polarizabilities
$(\alpha-\beta)_\pi$ shows agreement with well-known low-energy
theorems.
However, a discrepancy of almost a factor of 2 between this result
and several experiments remains, he said, so that more efforts are
required here.
In this respect, data from the COMPASS Collaboration at CERN may be
useful before further model-dependent interpretation is possible.

Hadron light-cone distribution amplitudes (LCDA) are the chief
ingredients of factorization formulae in QCD because they are
(i) process independent, i.e., universal, and
(ii) they encode in terms of appropriate expansion coefficients
the nonperturbative dynamics of confinement \cite{ER80,LB79},
while their evolution is controlled order-by-order by
renormalization-group type equations within QCD perturbation theory.
Evidently, their calculation/derivation is of paramount importance
for various phenomenological applications.
Several methods and models have been proposed over the years with
varying degree of precision and/or inherent theoretical uncertainties.
The interested reader may consult for further reading the reviews
in Refs.\ \cite{CZ84,BL89,Ste99}.

One of the most applied and most serious analytic approaches to
extract quantitative information on the hadron DAs is the method of
QCD Sum Rules---see \cite{CZ84}.
Pivovarov addressed in his talk the extraction of the kaon LCDA using
the QCD sum rule approach \cite{CKP07} and including NNLO perturbative
corrections.
He outlined that these perturbative corrections to the original
nonperturbative QCD sum rule are numerically important because they
change the relative magnitude of the $d=2$ loop diagrams and the
$d=4,6$ condensate terms in the Operator Product Expansion.
As a result, the first Gegenbauer moment $a_{1}^{K}$ of the
corresponding leading-twist kaon DA at a low scale $\mu \sim 1$~GeV
amounts to $a_1^K(1~\text{GeV})=0.10\pm 0.04$,
while the previous (average) result reads
$a_1^K(1~\text{GeV})=0.06\pm 0.03$.
There is, however, a rather large uncertainty in the determination of
$a_{1}^{K}$ owing to the poor precision of the light quark
masses---with $m_s$ directly entering the QCD sum rule---, whereas
$m_{u(d)}$ determine the quark-condensate densities via the
Gell-Mann--Oakes--Renner relation.
The comparison of the result for $a_1^K$ is also in disagreement with
very recent computations of this quantity on the lattice by the
QCDSF/UHQCD \cite{Bra06} and the UKQCD Collaboration \cite{Don07}.
They found, respectively,
$
 a_1^K(2 {\rm GeV})
=
 0.0453 \pm 0.0009 \pm 0.0029
$
and
$
 a_1^K(2 {\rm GeV})
=
 0.048 \pm 0.003 .
$
Two-loop evolution of Pivovarov's QCD sum-rule result to the
lattice scale $\mu^{\rm lat}=2$~GeV gives
$
 a_1^K(2 {\rm GeV})
=
 0.08 \pm 0.04
$,
which is---within the quoted uncertainties---only in marginal
agreement with the lattice estimates.
Clearly, here more work has to be done in order to understand the
roots of the observed discrepancy and achieve agreement between QCD
sum-rule calculations and lattice simulations.

The method of QCD sum rules underlies also the analysis of the
moments of the heavy-quark parton distribution function, reported
by Oganesian \cite{Oga07}.
The method was developed in \cite{BKO88} and is outlined in the first
entry of Ref.\ \cite{Oga07}.
The reliability of the heavy-quark mass limit of the sum rule (in the
description of the heavy-quark fragmentation functions), expressed in
terms of moments (calculated with QCD sum rules) of heavy-quark parton
distribution functions, was studied.
It was shown that in the case of the bottom quark, the
heavy-mass limit (of the expansion in the heavy-quark mass) is
reliable, provided the second $O(1/m^2)$ term is included.
In the case of the charm quark, the heavy-mass limit is not
reliable and, hence, the moments are far from the exact answer.
This implies, concluded Oganesian, that the heavy-quark limit is not a
reliable approximation for the parton distributions and fragmentation
functions of the $c$-quark.

Let us close this section by considering the study of the photon DA
and the role played by the magnetic susceptibility of the QCD vacuum
within the context of QCD sum rules with nonlocal condensates
\cite{MR89,BR91}, presented by Pimikov.
The final goal of his analysis is the extraction of the photon DA and
the vacuum magnetic susceptibility $\chi$ at $Q^2=0$ and
$\mu^2=1$~GeV${}^2$ including the NLO perturbative corrections.
This task has not yet been accomplished.
Hence, Pimikov restricted his report on the obtained results at the
LO level of the sum rules and concentrated on a nondiagonal correlator
to extract the photon DA and $\chi(0)$.
His main results are:
(i) A phenomenological estimate of
$\chi^{\text{Ph}}(0) = 4.05\pm 0.33~\text{GeV}^{-2}$.
(ii) The LO values
$\chi^{\text{LO}}(0)= 4.5\pm 0.5~\text{GeV}^{-2}$
and $\phi_\gamma^{\text{LO}}(x)=1$ (with the exception of the
endpoints).
He argued that the NLO magnetic susceptibility should be smaller, i.e.,
$\chi^{\text{NLO}}(0) = 4.0\pm 0.5~\text{GeV}^{-2}$, and emphasized
that the NLO terms should play a crucial role in the determination of
the photon DA in the endpoint regions.

\section{Hadron form factors}
\label{sec:form-factors}

Form factors constitute a powerful and effective tool in analyzing the
inner structure of hadrons and compare theoretical predictions with
experimental data with respect to charge and magnetic properties.
Coupled with QCD, hadron form factors provide a basic understanding of
the quark-gluon dynamics at the amplitude level.
Even more important, exclusive processes, naturally described in terms
of form factors, depend in detail on the composition of the hadron wave
functions themselves.
These are, as we have discussed earlier, the basic nonperturbative
ingredients in QCD calculations.
In view of the complexity of the QCD binding effects, it is not
possible to derive form factors from first principles of QCD.
Therefore, one attempts to model them as close to QCD as possible,
paying particular attention to factorization, causality, and
renormalization.
At low-momentum transfer, where hadronization effects become dominant,
form factors are combined with Vector Meson Dominance (VMD) constraints
in order to account for hadronic degrees of freedom.
In this way, it becomes possible to explore a wide region of $Q^2$ both
in the spacelike as well as in the timelike regime.

Naturally, the form factors of the nucleon are attracting much interest
both theoretically and phenomenologically.
From the theoretical point of view, one is keen to use them at large
$Q^2=-q^2$, where factorization applies, for different shapes of the
nucleon wave functions \cite{Ste99} and test whether they can provide
the correct sign, normalization, and scaling behavior as compared to
the data.
But also phenomenological approaches which combine the perturbative QCD
asymptotics with VMD at low $Q^2$ are indispensable in analyzing the
various existing data from SLAC and JLab, measuring the differential
cross section and polarization observables.

Tomasi-Gustafsson gave an overview on electromagnetic hadron form
factors, notably of the nucleon and the deuteron in the space- and
the timelike regions in terms of selected nucleon models
(see \cite{To-GuGa05,To-Gu06,To-GuGaRe06}).
She also presented model-independent features that are connecting
scattering (in Euclidean space) and annihilation (in Minkowski space)
channels, thus providing a deeper insight into the underlying physics.
Polarization experiments are especially crucial for the timelike
form factors as these are complex quantities.
In the presented work, parametrizations for the nucleon (and the
deuteron) form factors were used which work well in the spacelike
region, the motivation being provided by the fact that one can use
the abundance of the existing precise experimental data to constrain
the model parameters.
These parametrizations are then analytically continued to the
timelike region \cite{To-GuGa05}.
A particular role is played here by the ratio $F_2/F_1$ of the Pauli
over the Dirac form factor and its analytic continuation from the
spacelike to the timelike region.
Predictions were extracted for the magnetic and electric form factors
of the proton and the neutron in the Sacs parametrization and compared
in detail with experimental data from JLab.\footnote{For the
deuteron form factors, see also Adamu\v{s}\v{c}\'in's contribution in
the Proceedings and Dubni\v{c}ka's report to follow.}
An extension to spin-1 particles (and to the axial form factor of the
proton) was performed with the help of the VDM.
Predictions were presented for the differential cross section and for
polarization observables for the following processes:
$e^{+}e^{-}\to d+\overline{d}$,
$e^{+}e^{-}\to \rho+\rho\to 4\pi$,
and $e^{+}e^{-}\to a_1+\pi\to 4\pi$.
An important observation in the presented analysis is that the
asymptotic regime defined by the prescriptions of the considered models
and the asymptotic properties derived from the analyticity of the form
factors act at a different level.
In other words, the asymptotics of the total cross section for the $NN$
and $N\bar N$ interactions and the QCD asymptotics are connected to
each other in a non-trivial way.

In Dubnicka's talk, based on original work published in
\cite{Adamuscin05,Dubnicka04}, the size of the two-photon contribution
 to the elastic electron-deuteron scattering was estimated.
The new JLab proton polarization data on
$\mu_p G_{Ep}(t)/G_{Mp}(t)$ \cite{Jones_JLab00}
were analyzed together with all existing nucleon data on the
electromagnetic nucleon form factor in both the spacelike and the
timelike region (see \cite{Adamuscin05} for further references).
A ten-resonance unitary and analytic model was designed and used in
order to capture the key features of the electromagnetic structure
of the nucleon.
As a result, a non-dipole behavior of $G_{Ep}(t)$ in the spacelike
region with a zero around $t=-13$~GeV$^2$ was revealed.
The difference in the deuteron elastic structure functions
$A(t)$ and $B(t)$ between the non-dipole behavior of $G_{Ep}(t)$
and the Rosenbluth dipole behavior found to be negligible.
This outcome was interpreted as indicating that the two-photon-exchange
contribution to the unpolarized elastic $ed$-scattering is of less
importance relative to the one-photon exchange part.

In Dubni\v{c}kov\'a's talk a compatibility check of the new
spacelike-region pion's form-factor data with
$\sigma_{\rm tot}(e^{+}e^{-}\rightarrow$~hadrons),
obtained by radiative return, was outlined.
To this end, the new high-precision JLab data on the pion form factor
in the spacelike region \cite{Volmer_JLab00} were investigated as
regards their compatibility with the very precise data on the
total cross-section of the $e^{+}e^{-}$-annihilation process obtained
via radiative return by the KLOE Collaboration \cite{Aloisio04} and the
CMD-2 Collaboration \cite{Akhmetshin03}.
Technically speaking, this means to express the pion form factor in the
form of a dispersion relation and take into account its QCD asymptotic
properties.
Exploiting the analytic behavior of the pion form factor the way just
described, the compatibility of the two different experimental-data sets
was established.

\section{Quark confinement}
\label{sec:confin}

As mentioned earlier in this summary, a systematic treatment of QCD is
possible only within perturbation theory in the region of large
momenta, where the strong coupling is a small parameter of the
expansion.
In contrast, the region of low momenta, where quark and gluon degrees
of freedom start to hadronize into mesons and baryons, is not accessible
to perturbation theory and one has to rely upon effective QCD-inspired
models.
The deeper reason behind this failure, is the fact the there is no
any {\it ab initio} understanding of the mechanism that transforms
colored partons into colorless hadrons.
The technical name of this phenomenon is quark confinement and is
still the most puzzling issue (and stumbling block) of QCD as a
quantized Yang-Mills (YM) theory with local color gauge invariance.
Of course, one may hope that with increasing computer power, lattice
theory will provide a complete understanding of how quarks interact
with each other at large distances, revealing this way the confinement
mechanism (we have already some clues about this in terms of the string
tension).
However, one may strive to get an analytic understanding of confinement
and work out the crucial criteria for this in an unbroken YM theory,
like QCD.

Diakonov and Petrov dedicated strong efforts \cite{DP07,DP08} to
understand quark confinement in terms of dyons which are (anti-)self
dual solutions of the equations of motion of the pure YM theory.
In this work, presented by Petrov, two scenarios for confinement
at $T\neq0$ were addressed using a model based on dyons.
One scenario assumes that $<P>=0$ appears as the result of strong
fluctuations and the vacuum is essentially quantum.
The other option considers the state with $<P>=0$ to be the most
favorable one and emerging due to nonperturbative effects.
Fluctuations above this state are small and the situation is
semi-classical.
Skipping here technical details, let us concentrate on the main
findings of the presented approach:
(i) The semiclassical vacuum built of dyons has many features expected
for the confining pure YM theory.
(ii) It turns out that the minimum of the free energy for the dyon
ensemble lies exactly at the holonomy corresponding to the zero
Polyakov line $<P>=0$.
(iii) The dyon model reproduces all qualitative features of pure
gluon-dynamics known from lattice simulations.
(iv) Ideologically, it completely follows the t'Hooft--Polyakov
scenario of confinement and despite the fact that it is quite crude,
it appears to be numerically successful.

\section{Mathematical and other analyses}
\label{sec:math}

At the fundamental level, theorists are considering various ideas to
reveal the mechanism of the electroweak symmetry breaking.
In recent work \cite{CFN08} Faddeev and collaborators worked out a
version of the electroweak Lagrangian in terms of manifestly
gauge-invariant variables, which are the analogs of the Mei\ss ner
supercurrent that appears in the BCS superconductor.
These non-Abelian supercurrents remove all undesired gauge dependence
from the electroweak Lagrangian by a mere change of variables and
without any gauge fixing.
Then, as Faddeev pointed out, any issues with Elitzur's theorem become
obsolete.
Moreover, he showed that the ensuing Lagrangian describes the
electromagnetic interactions of the $SU_{L}(2)\times U_{\gamma}(1)$
gauge-invariant $W$ and $Z$ bosons and (overlooking topological
structures) it coincides with the original Lagrangian in the unitary
gauge.
Within this context, the Lagrangian acquires a generally covariant
form and the vector bosons receive their masses with no reference to
any symmetry breaking by a Higgs potential, provided the Higgs field
is interpreted as a dilaton that determines the conformal scale in a
locally conformally flat spacetime.

Let us complete this summary report by listing some more interesting
approaches that cannot be discussed here in detail, but may inspire the
interest of the readers.
Tkachev made a comment on the LLA method, the $k_T$ jet algorithm and
the BFKL theory.
Colferai studied a matrix formulation  for the small-$x$
renormalization-group improved evolution \cite{CCSS07}.
Krokhotin discussed BFKL effects in the jet production and a Monte
Carlo generator with BFKL evolution.
Was gave a comprehensive review of exact phase-space and spin
amplitudes and its applications in QCD NLO Monte Carlo programs.
Kojo addressed QCD Sum Rules and the $1/N_{c}$ expansion in connection
with dynamical studies of bare $2q$ and $4q$ poles in the $\sigma$
meson \cite{KJ08}.
Korchin presented a study of decays with light scalar mesons in
Resonance Chiral Perturbation Theory.
The electromagnetic form factors of kaons were computed \cite{IK06}
and compared with the data.
Buividovich reported on the entanglement entropy of gauge theories
and the holography for electric strings, worked out together with
Polikarpov \cite{BP08}.
Ktorides considered the AdS/CFT correspondence using Cartan's theory
of spinors.
Pajares analyzed rapidity long-range correlations, the color glass
condensate, and the percolation of strings \cite{ABP07,AMcLP06}.
Kudryavtsev presented a composite superstring model for hadrons,
employing the extended Virasoro superconformal symmetry \cite{Kud08}.
I apologize to those speakers whose contributions have been
omitted.

\section{Conclusions}
\label{sec:concl}

This conference took place just a couple of months before the start-up
of the LHC.
Many of the studies presented here may increase the precision of
theoretical calculations to the level relevant for the physics to be
probed at the LHC in the years to come.
To attain this goal, several issues, still at stake, have to be
clarified and improved techniques have to be further developed, like
the simultaneous resummation of QED and QCD large infrared effects.
Many speakers and authors of the papers presented here have given
efforts in advancing our understanding of these and other momentous
topics pertaining to a deeper understanding of QCD and its
applications.

\section*{Acknowledgements}
I would like to thank the organizers for giving me the honor
to summarize the theory talks of this interesting and topical workshop.
I am grateful to A.\ P.\ Bakulev for his invaluable help in preparing
the oral presentation of this summary report.
I am thankful to O.\ V.\ Teryaev for useful discussions and remarks,
and to A.\ P.\ Bakulev and S.\ V.\ Mikhailov for a critical reading of
the manuscript.
This work was partially supported by the Heisenberg-Landau
Program (Grant 2008) and the Deutsche Forschungsgemeinschaft under
contract 436RUS113/881/0.
A travel grant from the Sofja Kovalevskaja Preis der
Alexander von Humboldt Stiftung is gratefully acknowledged.

\end{document}